\documentclass[prb,showpacs,preprintnumbers,amsmath,amssymb,printfigures,twocolumn,superscriptaddress,10pt]{revtex4-1}
\usepackage[colorlinks=true,allcolors=blue]{hyperref}
\usepackage{graphicx}
\usepackage{amsmath}
\usepackage{bm}
\usepackage[T1]{fontenc}
\usepackage[latin9]{inputenc}
\usepackage{textcomp}
\usepackage{amstext}
\usepackage[Symbol]{upgreek}
\usepackage{booktabs}
\usepackage{dcolumn}
\usepackage{color}
\usepackage{enumerate}
\usepackage{latexsym,bm}
\usepackage{setspace}
\usepackage{braket}
\makeatletter

\begin{document}
\title{Tunable Band Gaps of In$_x$Ga$_{1-x}$N Alloys: From Bulk to Two-Dimensional Limit}

\author{V. Wang}
\thanks{wangvei@icloud.com.}
\affiliation{Department of Applied Physics, Xi'an University of Technology, Xi'an 710054, China}  

\author{Z. Q. Wu}
\affiliation{Department of Applied Physics, Xi'an University of Technology, Xi'an 710054, China}  

\author{Y. Kawazoe}
\affiliation{New Industry Creation Hatchery Center, Tohoku University, Sendai, Miyagi 980-8579, Japan} 

\author{W. T. Geng}
\thanks{geng@ustb.edu.cn.}
\affiliation{School of Materials Science \& Engineering, University of Science and Technology Beijing, Beijing 100083, China}

\date{\today}

\begin{abstract}
Using first-principles calculations combined with a semi-empirical van der Waals dispersion correction, we have investigated structural parameters, mixing enthalpies, and band gaps of buckled and planar few-layer In$_x$Ga$_{1-x}$N alloys. We predict that the free-standing buckled phases are less stable than the planar ones. However, with hydrogen passivation, the buckled In$_x$Ga$_{1-x}$N alloys become more favorable. Their band gaps can be tuned from 6 eV to 1 eV with preservation of direct band gap and well-defined Bloch character, making them promising candidate materials for future light-emitting applications. Unlike their bulk counterparts, the phase separation could be suppressed in these two-dimensional systems due to reduced geometrical constraints. In contrast, the disordered planar thin films undergo severe lattice distortion, nearly losing the Bloch character for valence bands; whereas the ordered planar ones maintain the Bloch character yet with the highest mixing enthalpies.       

\end{abstract}
\maketitle 

\section{introduction}

In the past few decades remarkable progress has been made in the development of optical and electronic devices based on group-III nitrides such as AlN, GaN, and InN as well as their alloys.\cite{Nakamura1994,Mishra2002,Yam2008} Especially, the band gap value in In$_x$Ga$_{1-x}$N alloys can vary from about 1.9 eV (InN) to 3.4 eV (GaN) with direct band gap character at room temperature, covering the emission spectral range from ultraviolet to near infrared. This makes In$_x$Ga$_{1-x}$N an excellent candidate material for light emitting diodes, laser diodes, and high-efficiency multi-junction solar cells.\cite{Orton1998,Wu2003} However, phase separation has been observed experimentally in a wide composition range,\cite{Singh1997,McCluskey1998,Kisielowski1997} due the large lattice mismatch 10\% between bulk InN and GaN\cite{Ho1996}. Recently, an enhanced alloy solubility and band-gap tunability in ternary InGaN nanowires were theoretically predicted by Xiang and co-workers.\cite{Xiang2008} Then an interesting question naturally arise: What will be different in alloy properties when In$_x$Ga$_{1-x}$N films are thinned to a few atomic layers?

Since the successful isolation of graphene,\cite{Novoselov2004,Novoselov2005,Schwierz2010} two dimensional (2D) materials such as hexagonal boron nitride,\cite{Novoselov2005a,Pacile2008} silicene,\cite{Sugiyama2010,Okamoto2010,Hare2012,Vogt2012} germanene,\cite{Bianco2013,Li2014b} stanene,\cite{Zhu2015} transition-metal dichalcogenides (TMDCs) with MX$_2$ composition (where M = Mo or W and X = S, Se or Te),\cite{Mak2010,RadisavljevicB2011} and phosphorene\cite{Li2014,Liu2014} have attracted tremendous attentions due to their novel electronic, optical, thermal, and mechanical properties for potential applications in various fields. The quantum confinement effect and reduced screening effect in 2D materials often manifest themselves in features clearly different from those of their bulk counterparts, and thus are under intensive sutdies.\cite{Song2012,Wang2012c,Zhang2005,Chhowalla2013,Xu2013,Butler2013,Zeng2015,Balendhran2015,Wang2015,Liu2015,Wang2015a}  A common feature of these 2D materials is that they are formed by stacking layers with strong in-plane bonds and weak, van der Waals (vdW) like interlayer attraction, allowing exfoliation into individual, atomically thin layers.

Very recently, Balushi \emph{et al.} have synthesized 2D GaN through migration-enhanced encapsulation growth technique.\cite{Balushi2016} The 2D GaN can be stabilized at ambient conditions with buckled structure when the surface dangling bonds are passivated. They reported that the band gap of buckled 2D GaN is thickness-dependent, varying from 3.4 eV in the bulk to 5.0 eV in a monolayer limit. By comparison, the band gap of buckled 2D InN is theoretically predicted to decrease from 3.0 eV in a monolayer to 1.9 eV in the bulk limit with a direct band gap. These results demonstrate that 2D nitrides are potential candidates for future tunable optoelectronics. Rubel \emph{et al.} performed first-principles calculations to investigate the alloying pf 2D planar-GaN with III-V substitutional elements and they predicted that that the buckled phase would be more attractive for optical emitters because of the intrinsic direct band gap character.\cite{Pashartis2017} 
  
The aim of this paper is to give a full understanding of the fundamental features in few-layer In$_x$Ga$_{1-x}$N alloys by performing first-principles calculations based on the density functional theory. Our calculations demonstrated that the mixing enthalpies buckled of In$_x$Ga$_{1-x}$N alloys decrease with decreasing layer thickness due to reduced geometrical constraints and negligible lattice distortion in the two-dimensional limit. An opposite trend is observed in planar 2D In$_x$Ga$_{1-x}$N alloys, a consequence of strengthened lattice distortion with decreasing layer thickness. As a result, the buckled alloys well preserve the well-defined Bloch character in both valence and conduction bands with a direct band gap which can be tuned from 6 eV to 1 eV by changing composition and layer thickness. These features make buckled films more attractive than planar ones for future light-emitting applications. The remainder of this paper is organized as follows. In Sec. II, methodology and computational details are described. Sec. III presents the numerical results of alloy formation enthalpy of In$_x$Ga$_{1-x}$N few-layers, followed by electronic structure analyses. Finally, a short summary is given in Sec. IV.

\section{Methodology}
\subsection{Computational details}
Our total energy and electronic structure calculations were performed using the Vienna Ab initio Simulation Package (VASP).\cite{Kresse1996, Kresse1996a} The electron-ion interaction was described using projector augmented wave (PAW) method \cite{PAW, Kresse1999} and the exchange and correlation (XC) were treated with generalized gradient approximation (GGA) in the Perdew Burke Ernzerhof (PBE) form\cite{Perdew1996}.
A cutoff energy of 400 eV was adopted for the plane wave basis set, yielding total energy convergence better than 1 meV/atom. In addition, the non-bonding van der Waals (vdW) interaction was incorporated by employing a semi-empirical correction scheme of Grimme's DFT-D2 method, which has been successful in describing the geometries of various layered materials.\cite{Grimme2006, Bucko2010} The semicore \emph{d} electrons of both Ga and In atoms were treated as core electrons. Test calculations show that the calculated lattice constant and band gap of GaN monolayer differ by less than 1.5\% from those of the corresponding configurations in which the \emph{d} electrons were included as valence electrons. The optimal lattice parameters are discussed in the next sections. The VASPKIT code was used to postprocess the VASP calculated data.\cite{vaspkit}

\begin{figure}[htbp]
\centering
\includegraphics[scale=0.58]{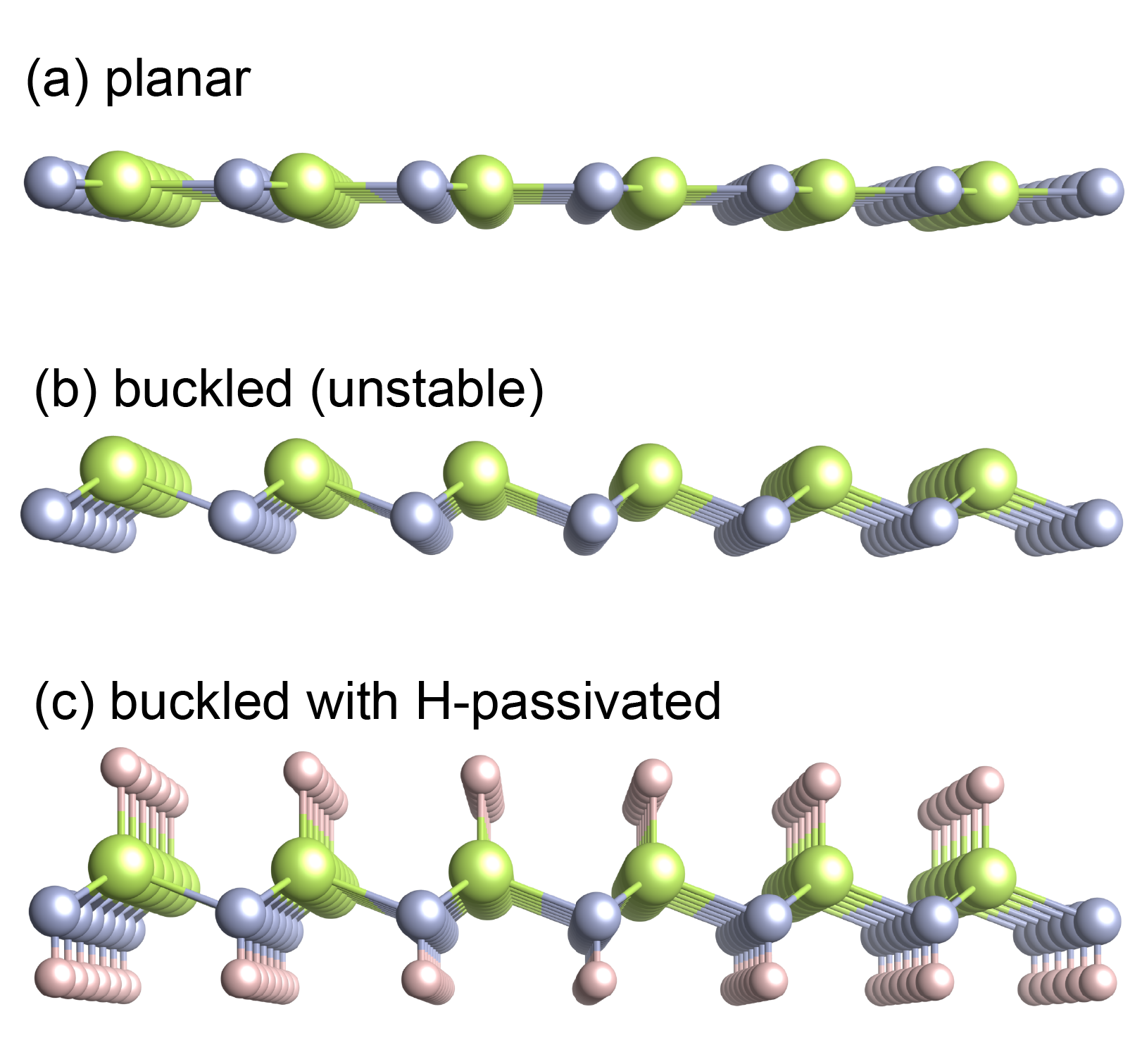}
\caption{\label{struct}(Color online) Ball-and-stick model for (a) planar, (b) buckled without H passivation and (c) buckled with H passivation GaN monolayer sheet. The green, silver and pink spheres refer to Ga, N and H atoms respectively.}
\end{figure}

In the slab model of few-layer GaN and InN systems, periodic slabs were separated by a vacuum layer of 15 {\AA} in \emph{c} direction to avoid mirror interactions. In sampling the Brillouin zone integrations, we used Monkhorst-Pack \emph{k}-point meshes with a reciprocal space resolution of 2$\pi$$\times$0.04 {\AA}$^{-1}$.\cite{Monkhorst1976} In geometry optimization, both the shapes and internal structural parameters of pristine unit-cells were fully relaxed until the residual force on each atom is less than 0.01 eV/{\AA}. The freestanding 2D GaN and InN are stable in planar structure, similar to graphene. This can be readily understood from the fact that the overlap between the Ga (In) $p_z$ and N-$p_z$ orbitals (to form $\pi$-bonds) could be maximized in planar structure. In addition, the Ga-N (In-N) bond length decrease from 2.01 (2.19) {\AA} in buckled monolayer to 1.88 (2.08) {\AA} in planar monolayer, and hence a further stabilized planar configuration. In contrast, the buckled films are more stable than the planar ones when the surface atoms are passivated with hydrogen, as illustrated in Fig. \ref{struct}. For comparison purposes, the tunable electronic structures of both planar and buckled phases of 2D In$_x$Ga$_{1-x}$N alloys are investigated in our current study.

Since the band gaps of semiconductors are often underestimated by conventional density functional theory (DFT) calculations with local or semilocal exchange-correlation functionals, part of electronic structure calculations were also performed using the Heyd-Scuseria-Ernzerhof (HSE06) hybrid functional\cite{Becke1993,Heyd2003,Perdew1996a,Krukau2006,Marsman2008} and quasiparticle GW methods\cite{Hedin1965, Shishkin2006,Paier2006,Fuchs2007,Shishkin2007} respectively for more rigorous scrutiny. It should be pointed that although the standard HSE06 approach can well yield the experimental band gap value in small-to-medium gap systems, it fails in wide-gap materials. Practically, the HSE06  can reproduce the experimental band gaps of elemental semiconductor by tuning the mixing- or/and screening-parameters empirically. However, this is not possible for alloys. Thus, the standard HSE06 approach was adopted in the following band structure calculations, namely the screening parameter $\mu$=0.2 {\AA}$^{-1}$ and the Hartree-Fock (HF) mixing parameter $\alpha$=25\% respectively.

The GW approach can generally attain better accuracy at the cost of higher computational effort as compared with hybrid DFT.  
To achieve good convergence of dielectric function in the G0W0 calculations, we used a large number of 80$\times$\emph{n} bands for each system, where \emph{n} is the total number of atom in the unit cell. The converged eigenvalues and wavefunctions, as well as the equilibrium geometry obtained from PBE functional, were chosen as the initial input for both G0W0 and HSE06 calculations. We note that the differences of lattice constants calculated by PBE and standard HSE06 respectively are less than 0.03 \AA. A 200-point frequency grid was applied to the integration over the frequencies along the imaginary time axis and real axis. Fuchs \emph{et al.} \cite{Fuchs2007}  pointed out that the one-shot G0W0 can provide a good agreement with experimental gap values. Note that in the G0W0 calculations only the quasiparticle energies were recalculated self-consistently in one iteration; the wave-functions were not updated but remained at the PBE or HSE06 level, henceforth denoted as G0W0@PBE and G0W0@HSE06 respectively. They also reported that G0W0@HSE06 has better performance than G0W0@PBE for materials consisting of $d$ electrons due to the improved description of the \emph{p-d} repulsion on the HSE06 level. 

\subsection{Alloy modeling}
The stability of 2D In$_x$Ga$_{1-x}$N alloys at 0 K can be evaluated from their mixing enthalpies ${\Delta}H_m$, which are calculated via
\begin{equation}\label{eq1}
\Delta H_{m}=E(\text{In$_{x}$}\text{G}\text{a}_{1-x}\text{N})-xE(\text{InN})-(1-x)E(\text{GaN}),
\end{equation}
where $E$(GaN), $E$(InN) and $E$(In$_x$Ga$_{1-x}$N) are the total energies of GaN, InN and mixed alloys. A negative value of ${\Delta}H_m$ implies that an ordered alloy can form spontaneously, while a positive one indicates tendency of phase segregation. 

In order to calculate the structural and electronic properties of In$_x$Ga$_{1-x}$N alloys, we adopted the special quasi-random structure (SQS) approach proposed by Zunger \emph{et al.}\cite{Zunger1990,Wei1990} The generation of a SQS supercell model was based on the lattice geometry, specifically the pair correlation function between lattice sites. A certain number of atomic configurations were created and the sum of pair correlation functions on each site was calculated. The distribution of atoms in a completely disordered (ordered) configuration has the minimum (maximum) relevant pair and multisite correlation functions. 
This approach is specially designed for small-unit-cell periodic structures (typically 2-16 atom/unit) that closely mimic the most relevant near-neighbor pair and multisite correlation functions of random alloys.
A distribution of local environments is conserved in SQS, the average of which corresponds to the random alloy, and thus the properties of the random alloy are well described. To investigate the bowing effect in random alloys, a 6$\times$6 supercell of few-layer GaN was used to study the effect of alloying with In concentration of 25\%, 50\% and 75\% via the SQS approach.

For the random substitutional alloys, the conventional band picture is broken due to the lack of translational symmetry. Nevertheless, the experimental data from alloys are very often interpreted in terms of such quantities as the effective mass, which can be readily derived from the band dispersion relations. Fortunately, the effective band structure (EBS) approach proposed by Zunger and co-workes can map the eigenvalues obtained from large supercell calculations into an effective band structure in the primitive cell using spectral function, and can recover an approximate $\emph{E}(\mathbf{k})$ for alloys. As a result, a direct comparison between the band structure from supercell calculations and that from primitive unit cell calculations become feasible. \cite{Popescu2010,Popescu2012} The spectral function is defined as,

\begin{equation}\label{sf}
A(\mathbf{k},E)=\underset{i}{\sum}|\braket{\mathbf{\Psi}_{i\mathbf{K}}|\mathbf{k}}|^{2}\delta(\varepsilon_{i}-E),
\end{equation}

where $\mathbf{k}$ is the wave vector in the first Brillouin zone of a primitive cell, and $i$ is the band index. $\mathbf{\Psi}_{i\mathbf{K}}$ and $\varepsilon_{i}$ are the eigenstate and eigenvalue of wave vector $\mathbf{K}$ in the supercell Brillouin zone respectively. For the perfect supercell of ordered systems, spectral function is either 1 or 0 because the symmetry does not change when transferring from the primitive cell to the supercell. However, in random alloys, the symmetry is broken, and different local environments in the alloys are reflected by arbitrary value between 1 and 0 of spectral function, resulting in a finite width in the EBS plots. 
In order to unfold the band structures of In$_x$Ga$_{1-x}$N alloys to the primitive cell representation, we performed an unfolding procedure using the BandUP code,\cite{Medeiros2014,Medeiros2015} a postprocessing interface to deal with the wave functions obtained by using VASP code.

The physical properties of many semiconductor alloys have a nonlinear dependence of the alloy composition $x$, such as lattice constants or band gaps. This is commonly described by a second-order polynomial of the form
 
\begin{equation}\label{bowing}
a_{\text{In}_x\text{Ga}_{1-x}\text{N}}=xa_\text{InN}+(1-x)a_\text{GaN}-bx(1-x),
\end{equation}

where $b$ is the so-called bowing coefficient which is a measure of how far the physical quantities deviates from the linear interpolation between pure phases. For isovalent alloys with a small chemical and size mismatch, the bowing coefficients are small constant numbers; for alloys with a large chemical and size mismatch, the bowing coefficients could be large and composition dependent. Usually, the bowing parameter of lattice constants is negligible, leading to a linear relationship, which is known as the Vegard's law.\cite{Vegard1921} On the other hand, the bowing of band gap and band edge position is often significant.

\section{Results and discussion}
\subsection{Pristine GaN and InN}
As a benchmark test, we have first investigated the fundamental properties of few-layer pristine GaN, InN and their bulk counterparts. Our calculated results as well as the available experimental data are listed in Tables \ref{buckled-table} and \ref{planar-table}. One can find that the formation energies are strongly thickness-dependent behavior. For example, the stability of the buckled GaN decreases by 0.4 eV when thinned from bulk to the monolayer limit. This suggests that the 2D nitride semiconductor becomes less stable with the decrease of film thickness. Furthermore, the buckled GaN (InN) monolayer is lower in energy by 0.87 (1.26) eV/formula unit than the corresponding planar configuration, and the similar trend is also found for the few-layer systems. It has been pointed out already the buckled GaN (InN) few-layers without H passivation, as shown in Fig. \ref{struct} (a) are unstable and will relax back to the corresponding planar ones. 

Although both buckled and planar InN few-layer compounds have positive formation energy, it is not necessarily true that they will decompose in a short time.\cite{Khachaturyan2013} The formation energy of a compound also depends on the growth conditions, such as external pressure, temperature, and strain. Recently, several experimental groups reported that a single-layer InN quantum well in a GaN matrix has been successfully synthesized by radio frequency plasma assisted molecular beam epitaxy (MBE).\cite{Yoshikawa2007,Dimakis2008,Pan2014} These InN/GaN multiple quantum wells are potentially applicable to room temperature operating excitonic devices working in near Ultraviolet colors. We find that the PBE-D2 predicted structural parameters are in good agreement with the experimental results for bulk GaN and InN. Nevertheless, the HSE06-D2 is more reliable than PBE-D2 on the formation energies for bulk GaN and InN. Interestingly, the lattice constants of 2D InN are more sensitive to the film thickness than those of GaN.  

We next investigated the role of H passivation in stabilizing the buckled configurations. We have calculated the adsorption energy of a H layer as a function of the number of GaN (InN) layers, in a similar procedure to what we have detailed in a previous work.\cite{Wang2012} For simplicity, we assume that two H layers are simultaneously adsorbed on both sides of the film. Then we can define the average adsorption energy $E_\text{H@XN}$=1/2($E_\text{H@X}$+$E_\text{H@N}$), where $E_\text{H@X}$ and $E_\text{H@N}$ are the adsorption energy of H on the top of X (X=Ga, In) and N atoms respectively. The $E_\text{H@XN}$ increases from -3.95  (-3.66) eV for a GaN (InN) trilayer to -3.22 (3.16) eV for a monolayer. Whereas the bond length $d_\text{H-Ga}$ (1.56 {\AA}), $d_\text{H-In}$ (1.76 {\AA}) and $d_\text{H-N}$ (1.03 {\AA}) nearly keep unchanged. We take buckled GaN few-layers as an example and analyze the contributions of $E_\text{H@Ga}$ and $E_\text{H@N}$ to $E_\text{H@GaN}$ by comparing partial density of states of Ga, H and N atoms (Supplementary Fig. \ref{figs1}). We find that the bonding states of H-$s$ and N-$p_z$ orbitals shift from -8 eV below Fermi level for a trilayer to -5 eV for a monolayer; while those of H-$s$ and Ga-$p_z$ orbitals are highly localized and expericence no shift. This means that the shift of these hybridized orbitals toward lower energy implies more stable bonding strength between H and N atoms. The above analysis tells us that the $E_\text{H@N}$ is mainly responsible for the tendency observed in $E_\text{H@GaN}$.

\begin{table*}[htbp]
\begin{ruledtabular}
\caption{\label{buckled-table} The calculated lattice constant \emph{a}, interlayer distance between two adjacent layers $\Delta$\emph{d}, formation enthalpy $E_f$ and band gap $E_g$ for buckled few-layer and bulk group-III nitrides using PBE, HSE06 and G0W0 approaches respectively.} 
\begin{tabular}{c|cc|cc|cccc|}
&\multicolumn{2}{c|}{Lattice (PBE calc.) }
&\multicolumn{2}{c|}{$E_f$ (eV/formula unit) }
&\multicolumn{4}{c|}{$E_g$ (eV)}\\
Systems & \emph{a} (\AA) &  $\Delta$\emph{d} (\AA) & PBE & HSE06 & PBE & HSE06  &  G0W0@PBE &  G0W0@HSE06   \\
\hline
monolayer GaN & 3.24  & -   & -0.86 & -1.04 & 3.21 &  4.38  & 5.95  & 5.99\\
bilayer GaN & 3.24 & 2.68   & -0.94 & -1.09 & 1.71 & 2.73  & 3.35  &  4.33  \\
trilayer GaN & 3.24  &  2.67  &  -1.04 &  -0.99  &  0.56  &  1.41  &  1.92   &    3.02 \\
bulk GaN &3.24 (3.19$^a$)  & 2.64 (2.59$^a$) & -1.23 & -1.38 (-1.29$^a$) & 1.68 & 2.88 & 3.42 & 4.67 (3.50$^a$)\\
monolayer InN   & 3.51  & -   & 0.40 & 0.04 & 2.32 &  3.51  & 4.90  & 5.45 \\
bilayer InN & 3.54 & 2.96  & 0.28 & -0.01 & 1.06 & 1.99  & 2.27  &  3.29 \\
trilayer InN & 3.54  &  2.95   & 0.28  & 0.37  & 0.23   & 0.88   & 0.87   &  1.58    \\
InN bulk &3.55 (3.54$^a$) & 2.87 (2.85$^a$) & -0.01 & -0.29 (-0.30$^a$) & metallic & 0.66 & metallic & 1.10 (0.70$^a$)  \\
\end{tabular} 
\leftline{$^a$ Experimental values in Ref. \onlinecite{Madelung2012,Lyons2014}.}
\end{ruledtabular}
\end{table*}

\begin{table*}[htbp]
\begin{ruledtabular}
\caption{\label{planar-table} The calculated lattice constant \emph{a}, interlayer distance between two adjacent layers $\Delta$\emph{d}, formation enthalpy $E_f$ and band gap $E_g$ for few-layer nitrides using PBE, HSE06 and G0W0 approaches respectively.} 
\begin{tabular}{c|cc|cc|cccc|}
&\multicolumn{2}{c|}{Lattice (PBE calc.) }
&\multicolumn{2}{c|}{$E_f$ (eV/formula unit) }
&\multicolumn{4}{c|}{$E_g$ (eV)}\\
Systems & \emph{a} (\AA) &  $\Delta$\emph{d} (\AA) & PBE & HSE06 & PBE & HSE06  &  G0W0@PBE &  G0W0@HSE06  \\
\hline
monolayer GaN  &3.25 & - &  0.01 & -0.02  &  1.95 &  3.37  &  4.15 & 5.12   \\
bilayer GaN  & 3.32  &  2.24  & -0.31  &  -0.47   & 1.82 & 3.30 & 3.68 &  4.71  \\
trilayer GaN  & 3.33   &  2.38   &  -0.44 &  -0.43    & 1.63   &   2.81  & 3.29   &  4.57   \\
monolayer InN &3.61 &  - & 1.66  &  1.71 &    0.50 &  1.69   &  1.62   &  3.01 \\
bilayer InN &  3.64 &  2.42  & 1.08   & 1.04  &  0.33 & 1.34   &  1.06  & 2.16  \\
trilayer InN  & 3.71  &  2.49   & 0.85   &  0.94    &  metallic   &  0.89   & metallic  &   1.02  \\
\end{tabular} 
\end{ruledtabular}
\end{table*}

\begin{figure}[htbp]
\centering
\includegraphics[scale=0.58]{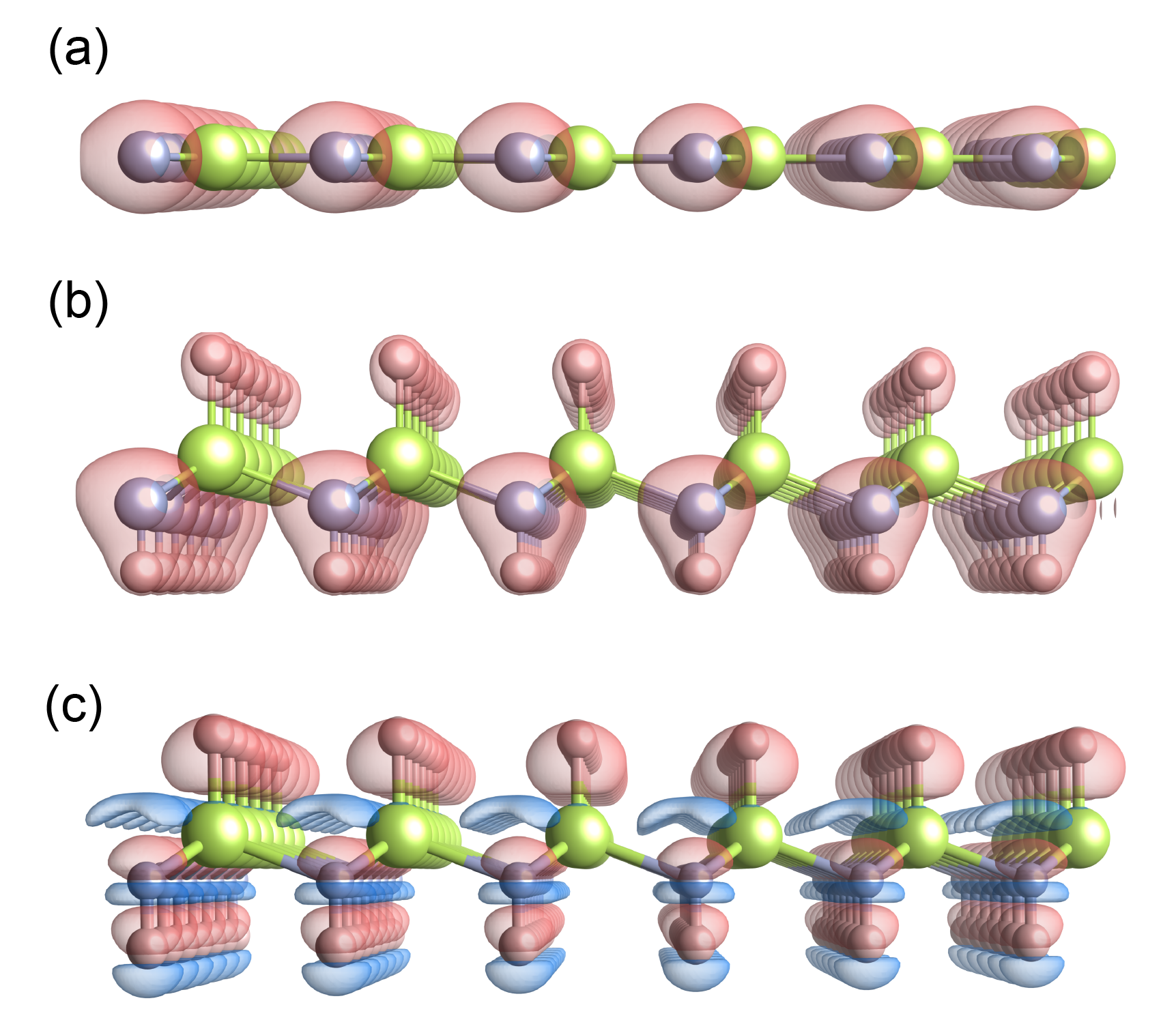}
\caption{\label{chg} (Color online) Charge density for (a) planar and (b) buckled GaN monolayer systems. (c) Charge density difference for buckled GaN monolayer. The regions of electron accumulation and depletion are displayed in pink and blue colors in (c), respectively.  The density isosurfaces are shown at 0.08 $e$/bohr$^3$. }
\end{figure}

From Figs. \ref{chg} (a) and (b) one can find that the charge densities of planar and buckled systems are concentrated around the N atoms, indicating that the Ga-N bonds are predominately ionic. To better perceive the H-N, H-Ga bonding, we display the change in charge density upon formation of chemical bonds, $\Delta\rho=\rho_\text{tot}-(\rho_\text{GaN}+\rho_\text{H})$, where $\rho_\text{tot}$, $\rho_\text{GaN}$ and $\rho_\text{H}$ are the total charge densities of buckled GaN monolayer with and without H adsorption, and  a hypothetical free-standing H layer. The regions of electron accumulation and depletion are displayed in yellow and blue colors, respectively. There is charge transfer from the Ga-N bonds to the H atom on top of Ga (H@Ga), indicating that the H@Ga atom behaves as an acceptor-like defect and the H-Ga bond is essentially ionic [see in Fig. \ref{chg} (c)]. By comparison, the charge accumulation in the regions between the H atom on top of H (H@N) and N atoms illustrates a covalent bond.

As for the band gap, we see from Tables \ref{buckled-table} and \ref{planar-table} that the PBE, HSE06 and G0W0@PBE approaches generally underestimate it for both GaN and InN in bulk form. It is noteworthy that the G0W0@PBE still predicts that InN is a metal, in sharp contrast to the observation of a 0.7 eV band gap.\cite{Wu2002} This can be understood from the fact that the GW algorithm implemented in VASP code is only intended for semiconductors and not well applicable to InN, which is predicted to be metallic at the PBE level. From the G0W0@HSE06 calculated  values we find that the band gap of bulk GaN and InN is overestimated by 1.2 eV and 0.4 eV respectively. Chen \emph{et al.} also observed similar trend in other wide-gap materials, especially for the systems with shallow \emph{d} bands such as GaN and ZnO. The noticeable discrepancies presumably come from the spurious \emph{p-d} hybridization introduced by the underestimation of 3\emph{d} binding energies in PBE calculations. By comparison, the hybrid functional lowers the localized 3\emph{d} bands owing to the reduced self-interaction error, thereby serving as a better starting point for these materials.\cite{Shih2010,chen2014} However, in this study we have treated Ga-3\emph{d} and In-4\emph{d} as core state and hence this could not explain the above discrepancy at the G0W0@PBE and G0W0@HSE06 levels.

\begin{figure*}[htbp]
\centering
\includegraphics[scale=0.75]{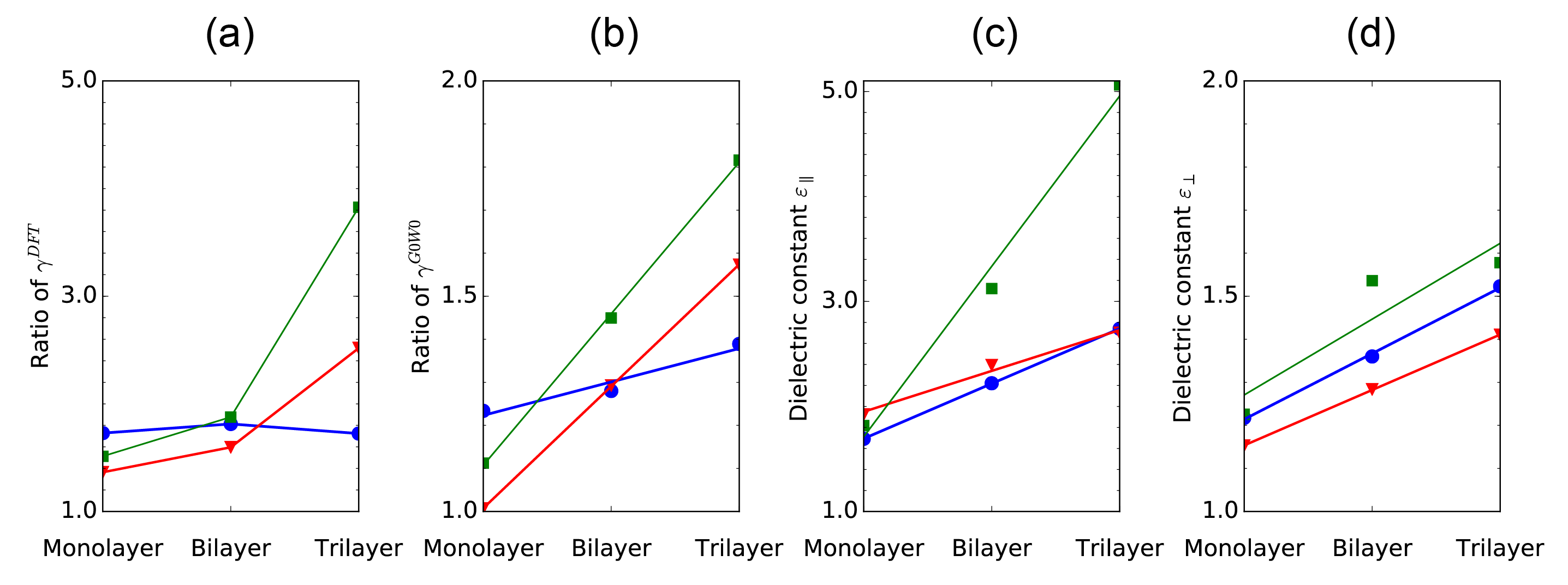}
\caption{\label{dielectric} (Color online) PBE-calculated ratios of gap magnitude (a) $\gamma^{DFT}$ and (b) $\gamma^{G0W0}$, static electronic dielectric constant along the (c) parallel direction $\varepsilon_{\parallel}$ and (d) perpendicular direction $\varepsilon_{\perp}$ to the layer. Red, green and blue lines represent the buckled GaN, InN and planar GaN few-layer systems respectively. }
\end{figure*}

It is reasonable to expect that the G0W0@PBE and G0W0@HSE06 give approximately the upper and lower bounds to the actual gap values of 2D GaN and InN fewlayers. An interesting finding is that the gap differences obtained by the G0W0@PBE and G0W0@HSE06 become more significant with increasing layer-thickness. To address this point, we define the relative difference gap magnitude as $\gamma^\text{DFT}$=$E_\text{g}^\text{HSE06}$/$E_\text{g}^\text{PBE}$ and $\gamma^\text{G0W0}$=$E_\text{g}^\text{G0W0@HSE06}$/$E_\text{g}^\text{G0W0@PBE}$, 
where $E_\text{g}^\text{HSE06}$, $E_\text{g}^\text{PBE}$, $E_\text{g}^\text{G0W0@HSE06}$ and $E_\text{g}^\text{G0W0@PBE}$ are the calculated band gap values obtained at the PBE, HSE06, G0W0@HSE06 and G0W0@PEB levels respectively. From Fig. \ref{dielectric} (a) one can find that the $\gamma^\text{DFT}$ shows a non-linear relationship with the layer number for all investigated systems. In contrast, the magnitude of $\gamma^\text{G0W0}$ increases linearly with the layer number. This seems to be a coincidence with the fact that the GW treatment is dependent on both static dielectric constant and interlayer separation.\cite{Hedin1965}  

To gain deeper insights into the linear scaling behavior of $\gamma^\text{G0W0}$, we also present the PBE-calculated static electronic dielectric constant of both in-plane $\varepsilon_{\parallel}$ and out-of-plane $\varepsilon_{\perp}$ as a function of layer number in Figs. \ref{dielectric} (c) and (d), respectively. The ionic contributions to the static dielectric constants are ignored because the ionic screening mainly takes effect during the geometry optimization. Interestingly, we find that the magnitude of both $\varepsilon_{\parallel}$ and $\varepsilon_{\perp}$ increase approximately linearly from monolayer to trilayer. A similar result has been obtained within HSE06 approach (not shown). In light of these findings, we conclude that the discrepancy between $E_\text{g}^\text{G0W0@HSE06}$ and $E_\text{g}^\text{G0W0@PBE}$ becomes more significant for the III-nitride semiconductors with higher electronic static dielectric constants. We do not expect it to be the case for other systems. It should be borne in mind that the dielectric function in 2D-insulators is not well defined and dependent on vacuum thickness.\cite{Cudazzo2011,Hueser2013} We take the buckled few-layer GaN as an example and revisit the dependence of dielectric function on the layer number with a fixed supercell thickness of 23 {\AA}. It turns out that both $\varepsilon_{\parallel}$ and $\varepsilon_{\perp}$ are weakly dependent on interlayer separation and remain to be a linear function of relationship with the layer number. However, the linear scaling behavior of $\gamma^\text{G0W0}$ with layer number is destroyed when a fixed interlayer distance is adopted (see Supplementary Fig. \ref{figs2}). 

\begin{figure*}[htbp]
\centering
\includegraphics[scale=0.55]{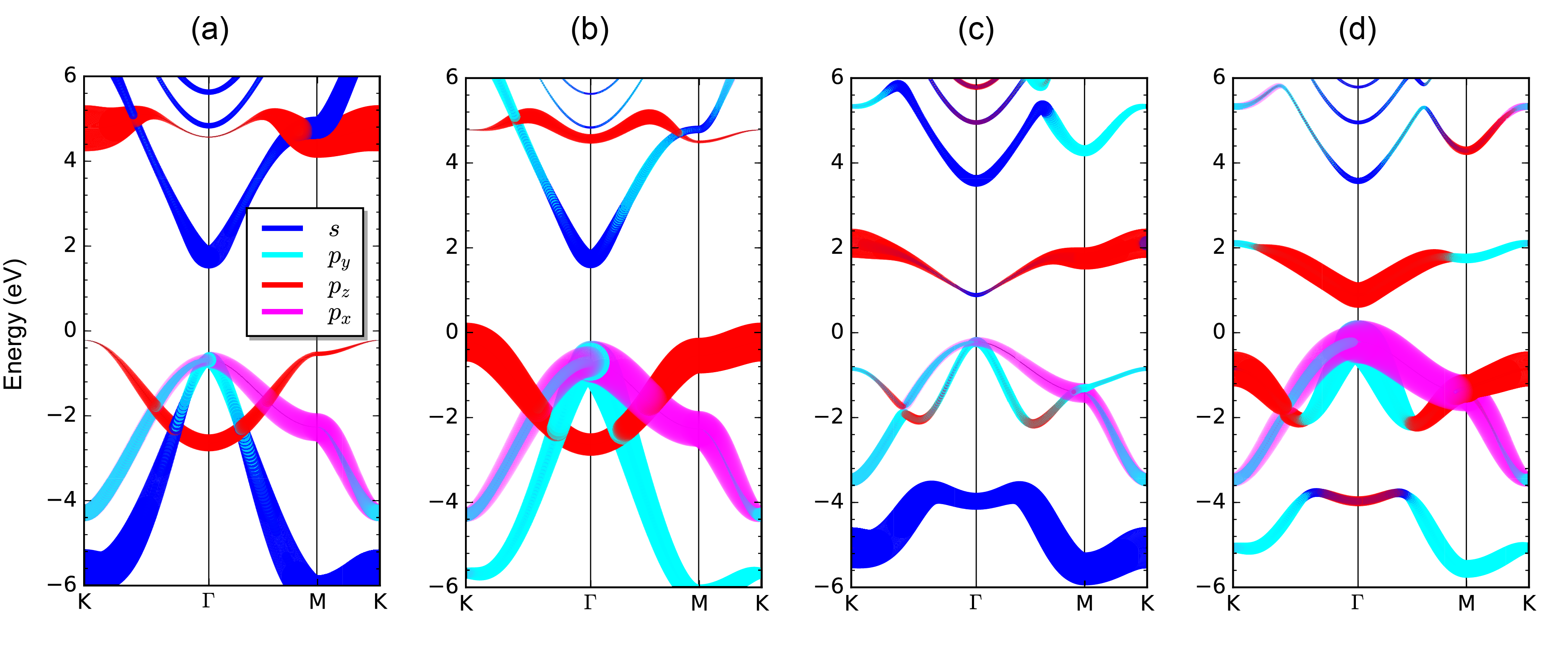}
\caption{\label{planar}(Color online) PBE-calculated orbital-projected band structure of (a) Ga, (b) N, (c) Ga and (d) N atoms in planar and buckled monolayer sheet without H passivation respectively. The line width indicates the weight of the component. The Fermi level is set to zero.}
\end{figure*}

\begin{figure*}[htbp]
\centering
\includegraphics[scale=0.55]{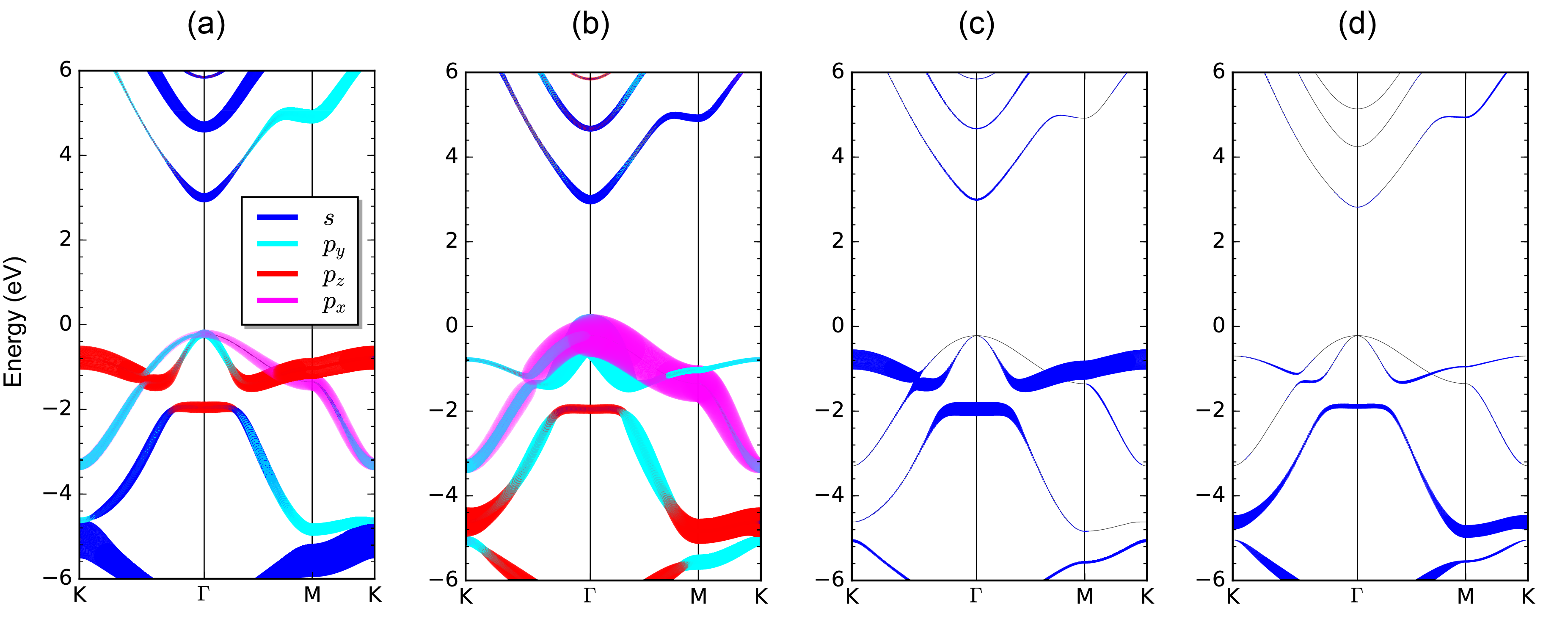}
\caption{\label{bukled}(Color online) PBE-calculated orbital-projected band structure of (a) Ga, (b) N, (c) H@Ga and (d) H@N atoms in buckled monolayer sheet with H passivation respectively. The line width indicates the weight of the component. The Fermi level is set to zero.}
\end{figure*}

To investigate the band characteristics of 2D GaN monolayers, the orbital-projected band structures of Ga, N and H atoms in the planar and buckled GaN monolayer sheets are displayed in Figs. \ref{planar} and \ref{bukled} respectively. The buckled configuration can be reached from a planar one through two steps: (i) the planar sheet wrinkles along the $c$ direction perpendicular to the GaN layer to form the buckled structure with unsaturated dangling bonds [see Fig. \ref{struct} (b)]; (ii), the H atoms adsorb on top of Ga and N atoms to passivate their dangling bonds. It can be seen from Figs. \ref{planar} (a) and (b) that the valence band maximum (VBM) of the planar structure occurs at the K point, and the conduction band minimum (CBM) at the $\Gamma$ point, resulting in an indirect band gap of 1.95 eV within PBE approach. The CBM mainly originates from the hybridization between Ga-$s$ and N-$s$ orbitals, while the VBM is composed mostly of the unsaturated N-$p_z$ orbital. From the planar sheet to a buckled one without H passivation,
the valance band dominated by N-$p_x$ and N-$p_y$  and conduction band dominated by N-$p_z$  at the $\Gamma$ point move toward the Fermi level, leading to a direct band gap of 0.95 eV at the PBE level, as displayed in Fig. \ref{planar} (d).

\begin{figure*}[htbp]
\centering
\includegraphics[scale=0.55]{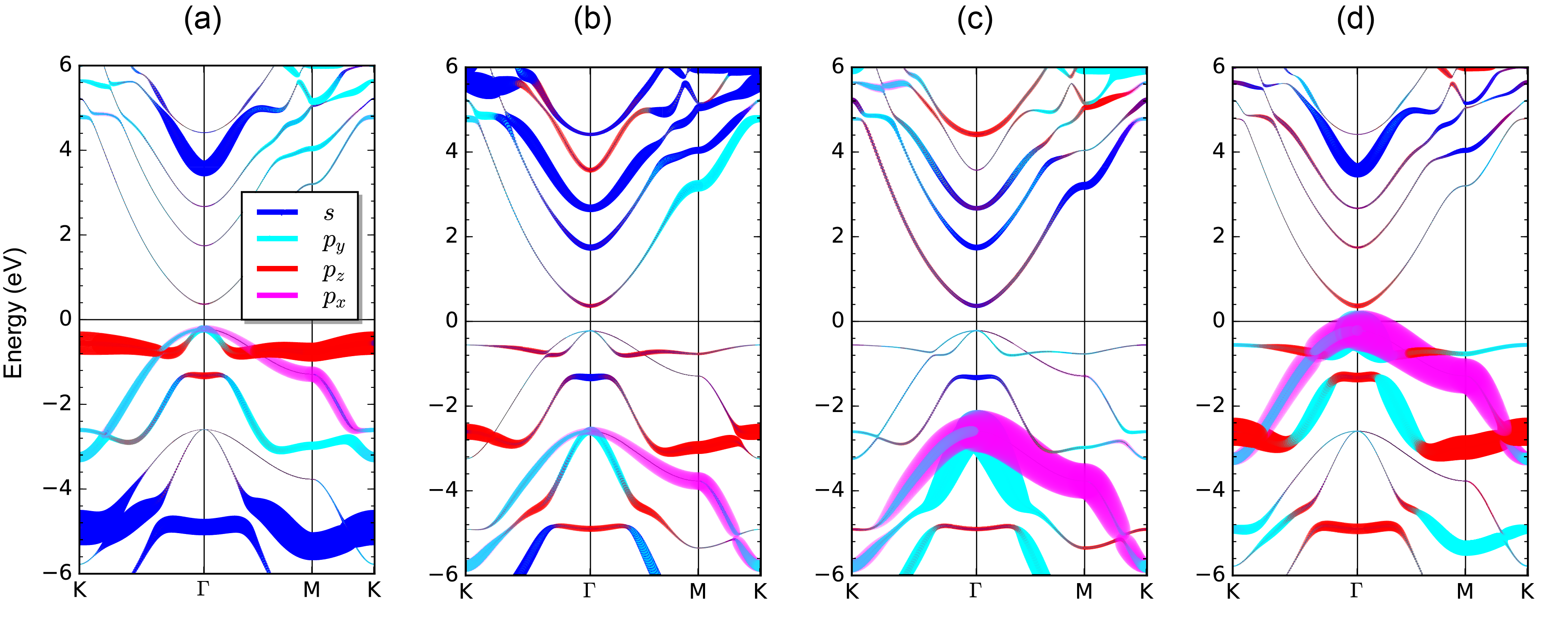}
\caption{\label{bilayer}(Color online) (Color online) PBE calculated orbital-projected band structure of (a) Ga$^\text{out}$, (b) Ga$^\text{in}$, (c) N$^\text{out}$ and (d) N$^\text{in}$ atoms in bilayer GaN. The  line width indicates the weight of the component. The Fermi level is set to zero.}
\end{figure*}

Next we turn to the buckled sheet with adsorbed H atoms. It is found that the adsorption of H atoms has little effect on the band dispersion around the $\Gamma$ point and the direct band gap is maintained. However, the conduction band with N-$s$ character is pushed to higher energy while the conduction band originating from the Ga-$s$-N-$s$ hybridization is shifted towards Fermi level and become the new CBM. As a result, the band gap is enlarged from the PBE value of 0.95 eV to 3.21 eV. On the other hand, the H atom (H@N) is covalently bonded with its neighboring N atom to form a $\sigma$ bond. It can be seen from Fig. \ref {bukled} (c) that the Ga-$p_z$ and H@Ga $s$-orbitals strongly hybridize in the energy range from 0 to -2 eV. The H@N $s$-states, on the other hand, are almost located at -2 eV below the Fermi level. In the unit cell of bilayer sheet, both Ga and N can reside in either outer or inner layers. The two inequivalent Ga (N) atoms are labelled as Ga$^\text{in}$ (N$^\text{in}$) and Ga$^\text{out}$ (N$^\text{out}$) respectively. As shown in Fig. \ref{bilayer}, the VBM of the bilayer sheet is dominated by the hybridization of $p_x$ and $p_y$ orbitals of N$^\text{in}$, with small contributions from the same orbitals of Ga$^\text{out}$ atom. While the CBM is composed of $s$ and $p_z$ orbitals from both N and Ga atoms. The band dispersions of the H-$s$ states in bilayer system are similar to the monolayer case but even more localized (Supplemental Fig. \ref{figs3}). As for the bulk system, its VBM is composed by equal proportion of $p_x$, $p_y$ and $p_z$ from N atom as a result of \emph{sp}$^3$-hybridization; while its band character of CBM resembles that in planar monolayer (Supplemental Fig. \ref{figs4}). The band character of InN is very similar to what has been observed for GaN (not shown here), albeit differ in the band gap. The HSE06-calculated band  dispersions show the same trends as the PBE ones but the absolute values differ significantly. Since the computational cost for GW calculations is very high, it is not feasible at present to carry out calculations of supercells with much more than 100 atoms within acceptable computation time. Thus, the standard HSE06 approach was adopted to predict the gap values of In$_x$Ga$_{1-x}$N alloys.

\subsection{In$_x$Ga$_{1-x}$N alloys}
 
\begin{figure*}[htbp]
\centering
\includegraphics[scale=0.7]{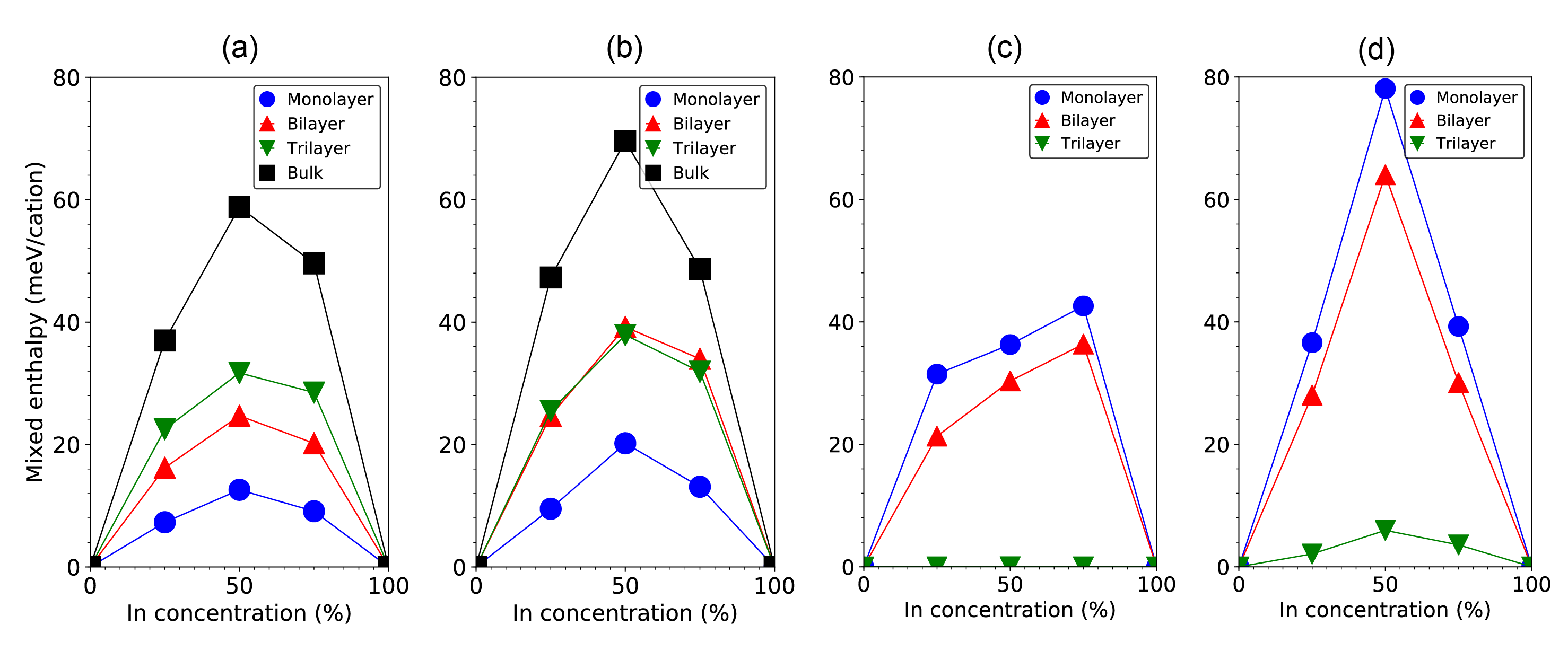}
\caption{\label{enthalpy} (Color online) PBE-calculated mixing-enthalpies [Eq. (\ref{eq1})] for buckled [(a) and (d)] and planar [(c) and (d)] Ga$_x$In$_{1-x}$N alloys in both disordered [(a) and (c)] and ordered [(b) and (d)] states. The blue circle, red triangle-down, green triangle-up, and black square symbols represent PBE calculated values for monolayer, bilayer, trilayer and bulk systems respectively. The corresponding solid lines are least-square fits of the data points to a quadratic polynomial. }
\end{figure*}

We present the PBE-calculated mixing enthalpy versus In concentration and layer thickness for random and ordered alloys in Fig. \ref{enthalpy}. It is found that the mixing enthalpy decreases monotonically when thinning buckled Ga$_x$In$_{1-x}$N alloys from bulk to monolayer. For example, at an In concentration of 50\%, the mixing enthalpy significantly decreases from 49.6 meV/cation for the bulk to 9.1 meV/cation for the monolayer. This suggests that the phase separation phenomenon observed in the experimental studies \cite{Singh1997,McCluskey1998,Kisielowski1997} could be suppressed in the two-dimensional limit, especially at 25 \% In concentration. These results can be easily understood from the fact that fewer geometrical constraints in low dimensional than in their bulk counterparts. Moreover, the mixing enthalpies [Figs. \ref{enthalpy} (b) and (d)] show a parabolic-like concentration dependence except for the disordered planar alloys which experience severe lattice distortion. A more detailed discussion will be given in the following section. At a finite temperature \emph{T}, the stability of random alloys can be described by the Gibbs free energy $G(x, T)$ which consists of mixing enthalpy and entropy contributions, defined as
\begin{equation}\label{freeenergy}
G(x,T)=\Delta H_m(x)-TS(x),
\end{equation}
where $\Delta H_m(x)$ is the mixing enthalpy defined in Eq. (\ref{eq1}). $S(x)$ is the mixing entropy which can be estimated based on a mean-field approximation,\cite{Lambrecht1993,Alling2007} 

\begin{equation}\label{entropy}
S(x)=-k_B[x\text{ln}x+(1-x)\text{ln}(1-x)],
\end{equation}
where $x$ and $k_B$ are the concentration of solute and the Boltzman constant respectively. Clearly, the inclusion of contribution from the mixing entropy can further enhance the solubility of In in GaN monolayer by increasing the temperature. Note that the mixing enthalpies of ordered systems are higher than the corresponding disordered ones in the absence of entropic contribution. This is because the disordered configurations include more or less local composition segregation. In sharp contrast, an opposite tendency is observed in planar Ga$_x$In$_{1-x}$N alloys, \emph{i.e.}, the mixing enthalpy increases with layer number. 

To gain deeper insights into the stability of few-layer Ga$_x$In$_{1-x}$N alloys, the formation process of $x$ GaN + ($1-x$) InN $  \rightarrow$ Ga$_x$In$_{1-x}$N is divided into three steps: First, the lattice of a$_\text{GaN}$ (a$_\text{InN}$) is expanded (compressed) to the equilibrium lattice a$_\text{Ga$_{x}$In$_{1-x}$N}$ of the intermediate compound, accompanied by partial enthalpy contribution $\Delta E^\text{VD}$ due to the uniform elastic volume deformation. Second, both (1-\emph{x}) GaN and \emph{x} InN are prepared at the lattice value of a$_\text{Ga$_{x}$In$_{1-x}$N}$ and brought together to form the crystal Ga$_{x}$In$_{1-x}$N with all atoms at their ideal lattice positions. The chemical electronegativity exchange $\Delta E^\text{CE}$ represents the ability of exchange charge in the combined system. And in the final step the internal structural degrees of freedom are relaxed to 
reach equilibrium structure, releasing a structural energy $\Delta E^\text{ST}$.\cite{Srivastava1985} With the definition above, we can obtain $\Delta H_m(x)$=$\Delta E^\text{VD}$+$\Delta E^\text{CE}$+$\Delta E^\text{ST}$. 

\begin{figure*}[htbp]
\centering
\includegraphics[scale=0.98]{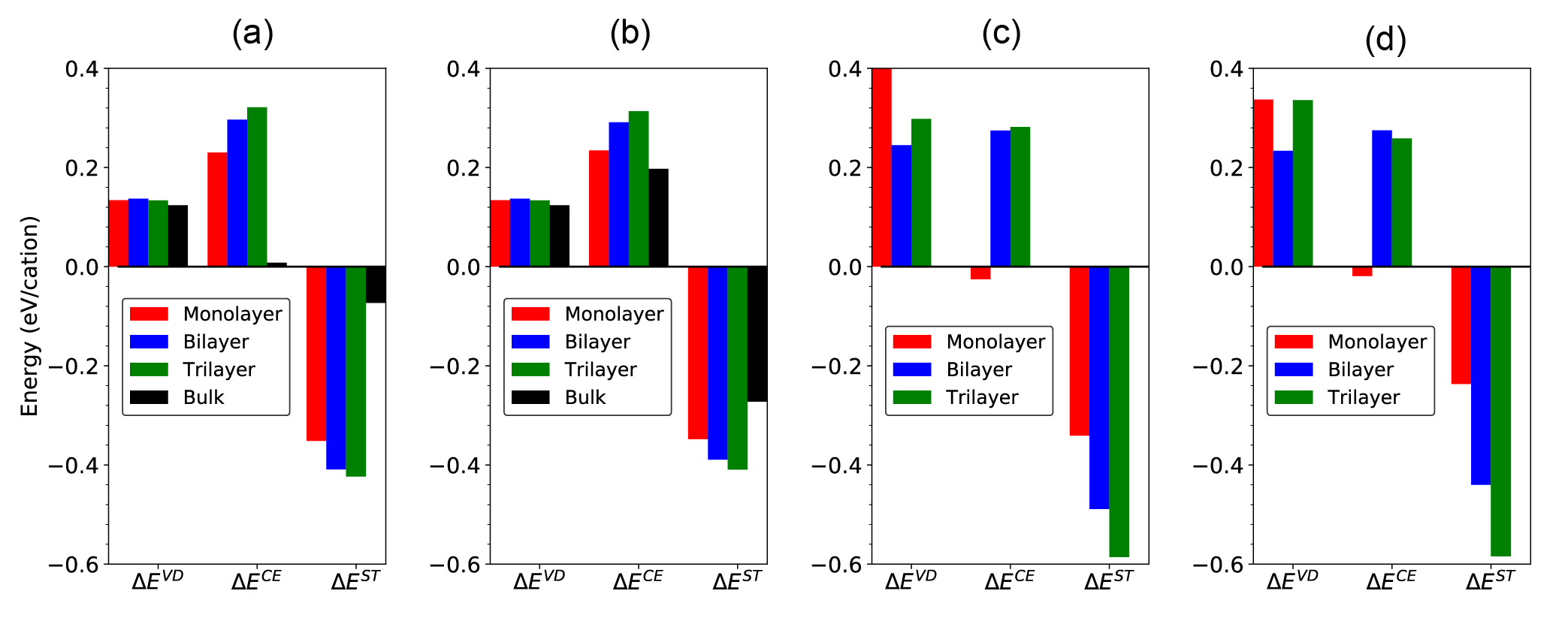}
\caption{\label{enthalpy_split} (Color online) PBE-calculated volume deformation $\Delta E^\text{VD}$, chemical electronegativity exchange $\Delta E^\text{CE}$, structure relaxed $\Delta E^\text{ST}$ contributions to mixing enthalpy in (a) buckled disordered and (b) buckled ordered, (c) planar disordered and (4) planar ordered  In$_{0.5}$Ga$_{0.5}$N alloys. }
\end{figure*}

We take Ga$_{0.5}$In$_{0.5}$N alloys as an example and present the calculated results of these three contributions to mixing enthalpy in Fig. \ref{enthalpy_split}. We find that the $\Delta E^\text{VD}$ is around 0.13 eV/cation and almost independent to the layer thickness in buckled alloys. However, the magnitude of $\Delta E^\text{VD}$ in planar configuration is larger than 0.25 eV/cation and varies strongly with layer number. This implies that the stability of planar GaN and InN is more sensitive than those of the buckled ones on the changes of lattice constants, in agreement with the trend observed in the lattice changes of planar configuration versus the number layer of listed Table \ref{planar-table}. Moreover, the chemical electronegativity exchange $\Delta E^\text{CE}$ has a positive value, indicating accumulation of extra charge on the weaker bond, and becomes less stable when changing from monolayer to trilayer. The only exception is the planar monolayer which has a negative value of $\Delta E^\text{CE}$ ($\sim$20 meV/cation), suggesting GaN (more ionic bond) is more stable than InN (less ionic bond).\cite{Srivastava1985} The magnitude of $\Delta E^\text{ST}$ is associated with lattice distortion. That is the higher value of $\Delta E^\text{ST}$ means the more significant distortion. Clearly, the severest deformation is observed in the planar disordered Ga$_x$In$_{1-x}$N few-layer alloys, followed by the corresponding ordered ones, while all the buckled alloys have the weakest deformation and are less sensitive to the layer number. 

The resulting relaxed geometries of Ga$_{0.5}$In$_{0.5}$N monolayer and bilayer are displayed in Figs. \ref{figs5} and \ref{figs6} of Supplementary Materials. 
As expected, buckled alloys undergo moderate lattice distortion. Whereas the relaxed structures of planar alloys become completely wrinkled, partly due to large lattice mismatch of $\sim$10\%. Furthermore, the weak $\pi$ bonds in the two outermost layers are also responsible for the instability of planar alloys. Such a significant lattice-distortion could affect the band dispersion characteristics in planar alloys, as will be discussed later. The ordered planar alloys remained planar upon structural relaxation. The is because that the strain-induced effects is partly canceled out in a uniform distribution case.

\begin{figure*}[htbp]
\centering
\includegraphics[scale=0.8]{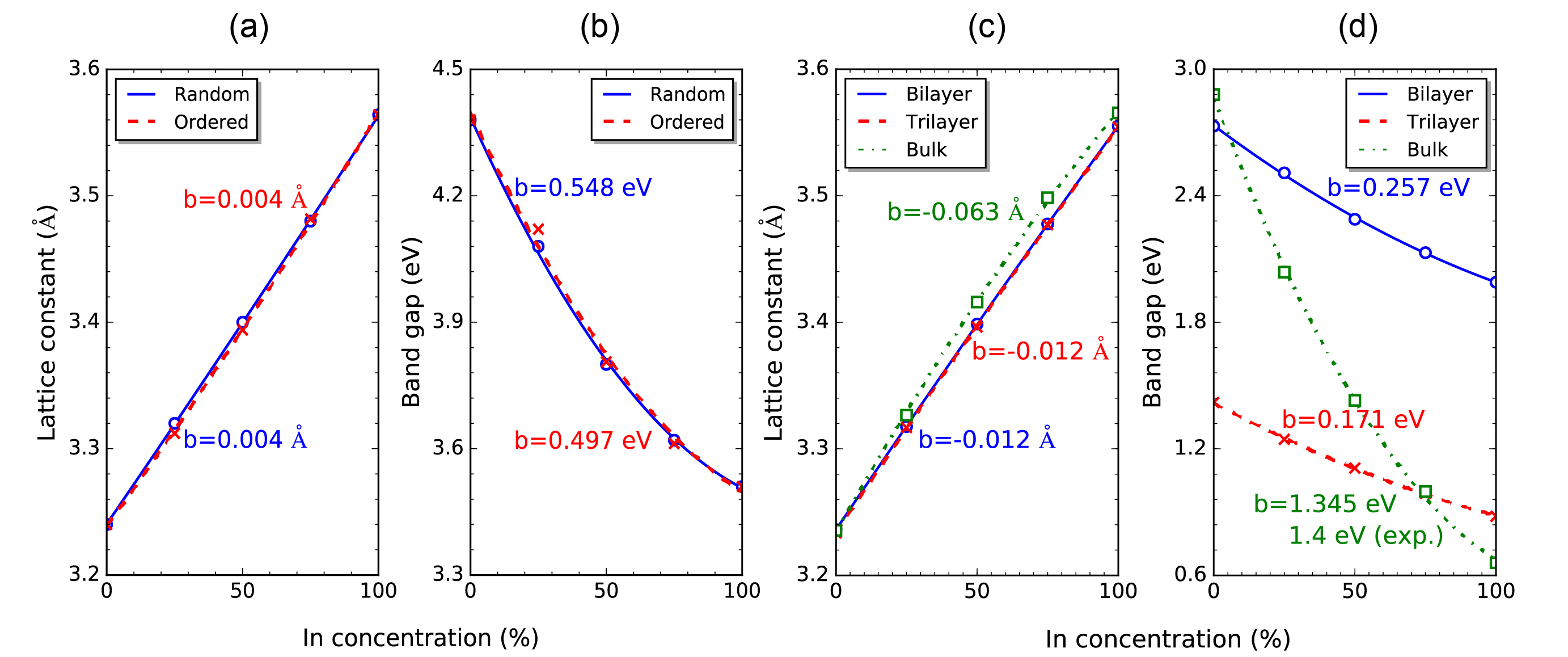}
\caption{\label{bowing-buckled}(Color online) (Color online) (a) and (c) PBE-calculated Lattice constant $a$ and (b) and (d) HSE06-calculated band gap $E_g$ versus In concentration $x$ in few-layer Ga$_x$In$_{1-x}$N. The blue and red lines are the parabolic fit based on Eq. \ref{bowing} for random and ordered configurations respectively.}
\end{figure*}

\begin{figure*}[htbp]
\centering
\includegraphics[scale=0.8]{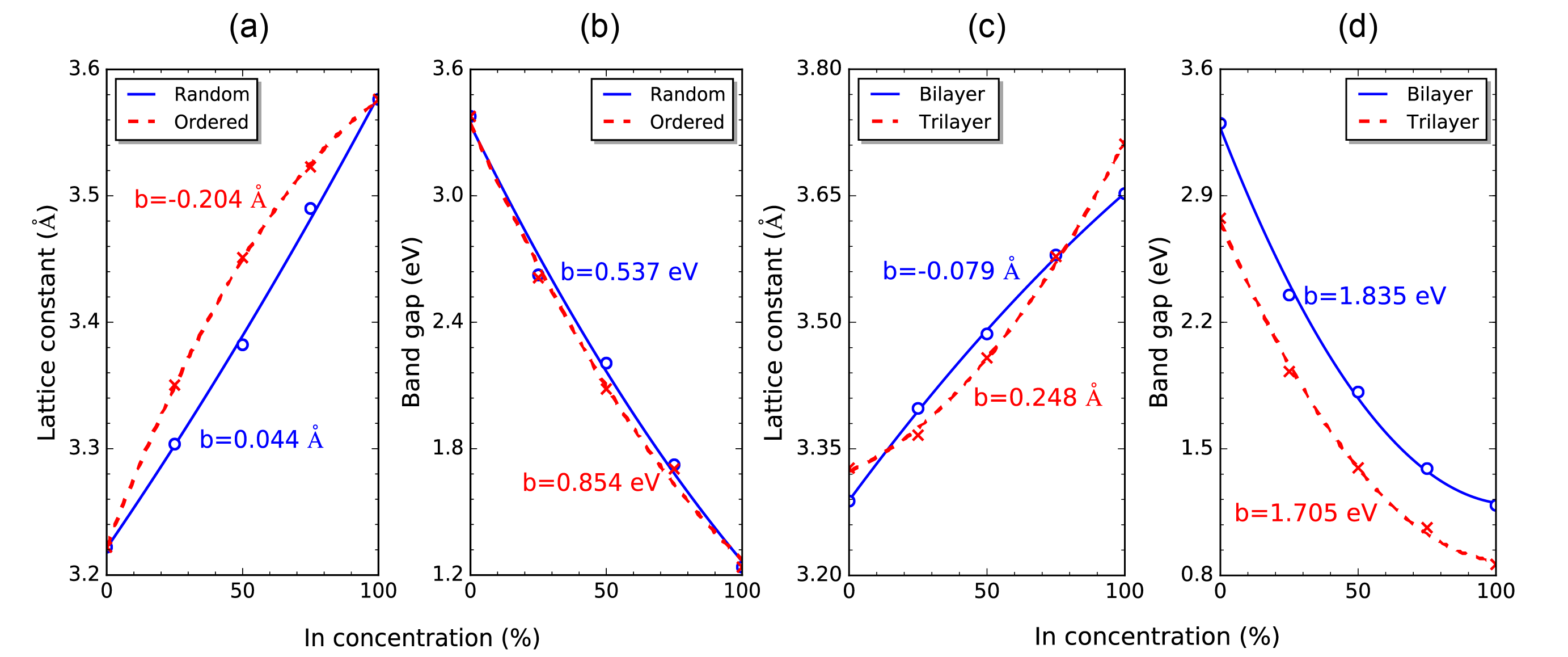}
\caption{\label{bowing-planar}(Color online) (Color online) (a) and (c) PBE-calculated Lattice constant $a$ and (b) and (d) HSE06-calculated band gap $E_g$ versus In concentration $x$ in few layer Ga$_x$In$_{1-x}$N. The blue and red lines are the parabolic fit based on Eq. \ref{bowing} for random and ordered configurations respectively.}
\end{figure*}

We plot in figures \ref{bowing-buckled} (a) and (c) the lattice constant $a(x)$ of Ga$_x$In$_{1-x}$N versus alloy composition $x$. We can see $a(x)$ increase linearly with increasing In content in accordance with Vegard's Law, with a negligible bowing parameter \emph{b} less than 0.01 {\AA} for buckled phase. However a slight deviation on the order of 0.2 {\AA} is observed in the planar one. The band gap energy exhibits a notably nonlinear change with alloy composition, and the average bowing parameter \emph{b} is as large as 0.58 eV. Despite the nonlinearity, the band gap energy changes continuously with varying alloy composition and can be purposefully tuned to any intermediate value by adjusting the In to Ga ratio. Furthermore, the bowing parameters of both lattice constant and band gap are found be insensitive to the distribution of alloy constituents. On the other hand, the bowing parameters of both lattice constant and band gap for planar alloys vary significantly with the layer number and the ordering of alloy constituents, as shown in Fig. \ref{bowing-planar}. We predict a band gap bowing value \emph{b}=1.345 eV for bulk Ga$_x$In$_{1-x}$N alloy, which is in good agreement with the available experimental value of 1.4 eV.\cite{Vurgaftman2003} Nevertheless, Moses \emph{et al.} reported that a single bowing parameter cannot accurately describe nonlinearities over the entire composition range by performing hybrid-functional calculations.\cite{Moses2011}

\begin{figure*}[htbp]
\centering
\includegraphics[scale=0.6]{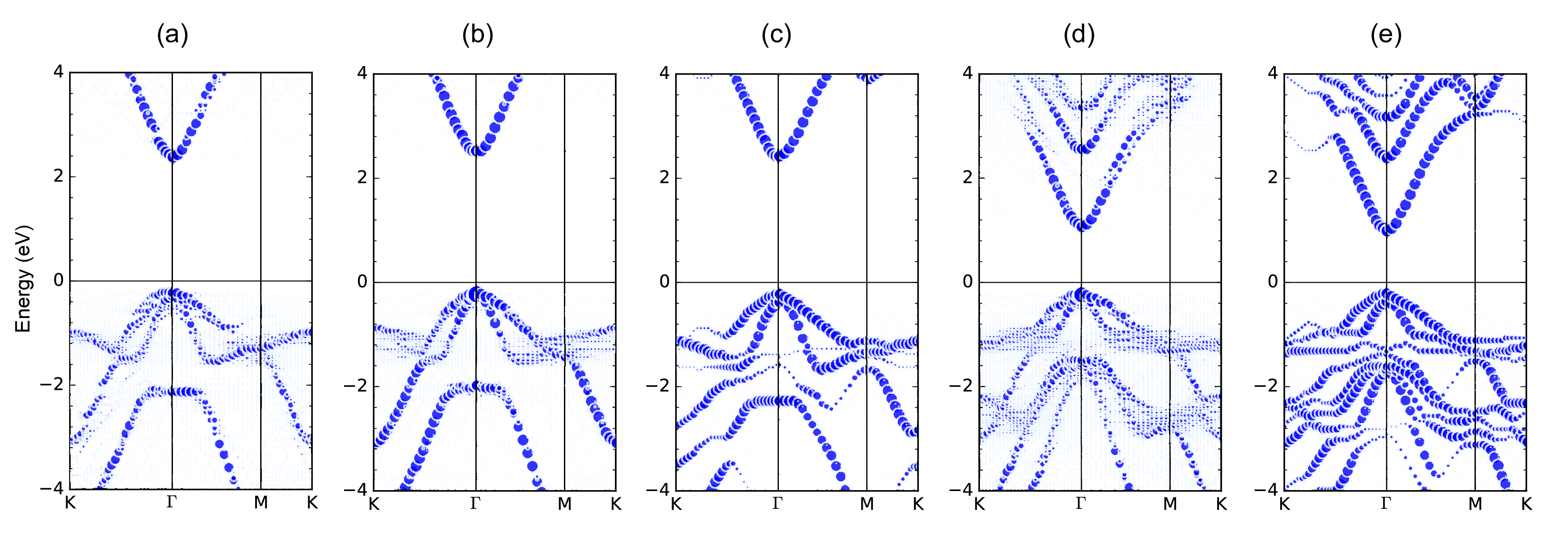}
\caption{\label{ebs-buckled}(Color online) PBE-calculated effective band structure of buckled In$_{0.5}$Ga$_{0.5}$N alloys derived from a 6$\times$6 supercell unfolded to a primitive Bloch representation. (a) disordered and (b) ordered monolayer configurations,  (c) disordered monolayer without atomic positions relaxation, (d) disordered and (e) ordered bilayer configurations.}
\end{figure*}

\begin{figure*}[htbp]
\centering
\includegraphics[scale=0.6]{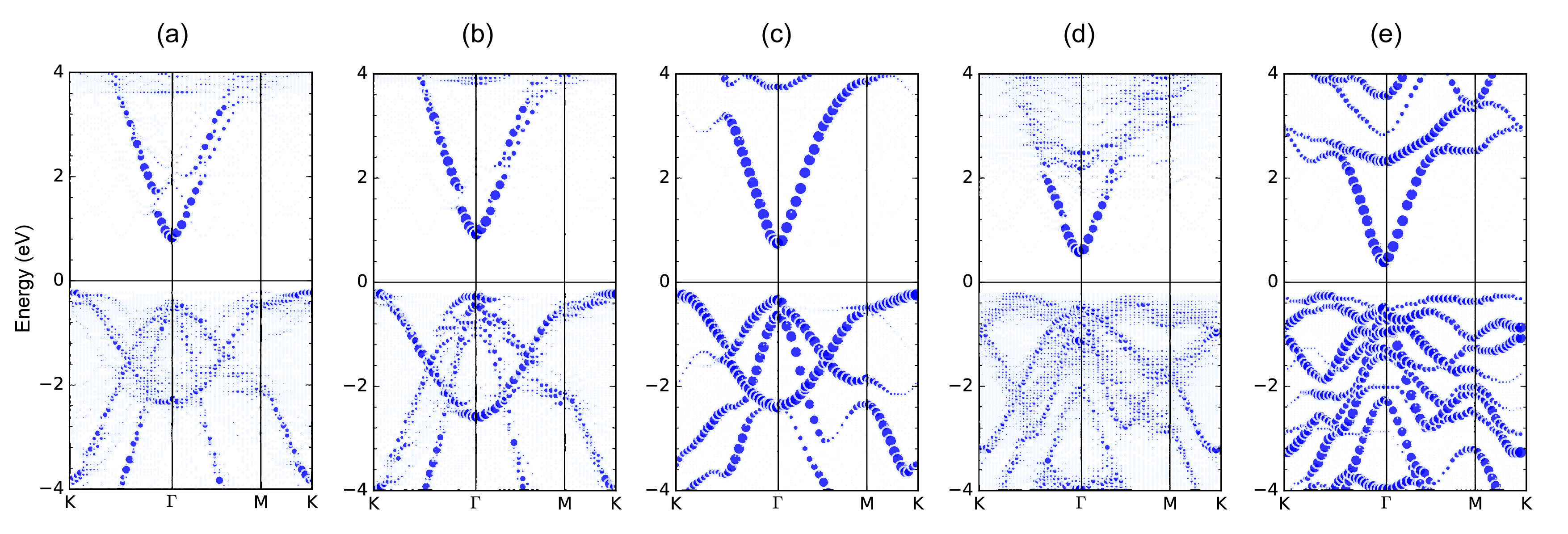}
\caption{\label{ebs-planar}(Color online) PBE-calculated effective band structure of planar In$_{0.5}$Ga$_{0.5}$N alloys derived from a 6$\times$6 supercell unfolded to a primitive Bloch representation. (a) disordered and (b) ordered monolayer configurations,  (c) disordered monolayer without atomic positions relaxation, (d) disordered and (e) ordered bilayer configurations.}
\end{figure*}

We present the PBE-calculated EBSs for buckled and planar few-layer alloys in Figs. \ref{ebs-buckled} and \ref{ebs-planar} respectively. Among these systems, both buckled monolayer and bilayer with ordered constituents maintain well the Bloch characters in a wide energy range [-4, 4] eV near the Fermi level due to the ordered chemical effects and small lattice distortions, as shown in Figs. \ref{ebs-buckled} (b) and (d). The band dispersion characters near the CBM around the $\Gamma$ point for the buckled disordered configurations are overall less disturbed when compared with the band structures of the pristine systems. The Bloch characters near VBM, however, are seriously weakened. To investigate the effect of lattice distortions on the Bloch character, we take the disordered monolayer as an example and plot the EBS of this system without atomic positions relaxation [Fig. \ref{ebs-buckled} (c)], that is, all atoms are constrained to their ideal positions at the equilibrium lattice. Clearly, the strain effect is mainly responsible for the partial loss of the Bloch character observed in disordered buckled-alloys. The valence bands near the Fermi level nearly disappear in disordered planar-alloys due to the sever lattice distortions [Figs. \ref{ebs-planar} (b) and (d)]. In addition, the planar-alloys still maintain the indirect nature of their band gaps. These two factors could limit the potential of planar few-layer alloys for applications in the next generation of tunable nanoelectronic and optoelectronic devices. 

\section{summary}
To explore the fundamental properties of few-layer In$_x$Ga$_{1-x}$N compounds, we have performed first-principles calculations based on the density functional theory to study the structural parameters, mixing enthalpies, and band gaps of buckled and planar few-layer In$_x$Ga$_{1-x}$N thin films. A semi-empirical van der Waals dispersion correction was employed to describe the non-bonding interlayer interactions. We provide computational evidence that the free-standing buckled films are less stable than the planar ones. With hydrogen passivation, nevertheless, the buckled In$_x$Ga$_{1-x}$N configurations become more favorable, with tunable band gaps ranging from 6 eV to 1 eV. Both the direct gap and well-defined Bloch character are preserved for the energy bands, making them promising candidate materials for future light-emitting applications. Because of reduced geometrical constraints, phase separation could probably be suppressed in these two-dimensional systems. We find that the structural and electronic properties of buckled thin films do not depend sensitively on the indium distribution. The planar thin films with disordered indium suffer severe strains and the Bloch character of the valence bands nearly fades away. By comparison, the planar thin films with ordered indium maintain the Bloch character, albeit with the high mixing enthalpies which make them unstable. 
\begin{acknowledgments}
This work is supported by the support of Natural Science Basic Research Plan (Program No. 2017JM1008) and Natural Science Foundation (Grant No.2014JM2-5049) of Shaanxi Province of China. The calculations were performed on the HITACHI SR16000 supercomputer at the Institute for Materials Research of Tohoku University, Japan. The authors thank K.Tanno and N. Igarashi provide technical support for high performance computing resources.
\end{acknowledgments}
\nocite{*}
\bibliographystyle{aipnum4-1}
%

\pagebreak
\widetext
\begin{center}
\textbf{\large Supplemental Materials: \\Tunable Band Gaps of In$_x$Ga$_{1-x}$N Alloys: From Bulk to Two-Dimensional Limit}
\end{center}
\setcounter{equation}{0}
\setcounter{figure}{0}
\setcounter{table}{0}
\setcounter{page}{1}
\makeatletter
\renewcommand{\theequation}{S\arabic{equation}}
\renewcommand{\thefigure}{S\arabic{figure}}
\renewcommand{\bibnumfmt}[1]{[S#1]}
\renewcommand{\citenumfont}[1]{S#1}

\begin{figure}[htbp]
\includegraphics[scale=1.2]{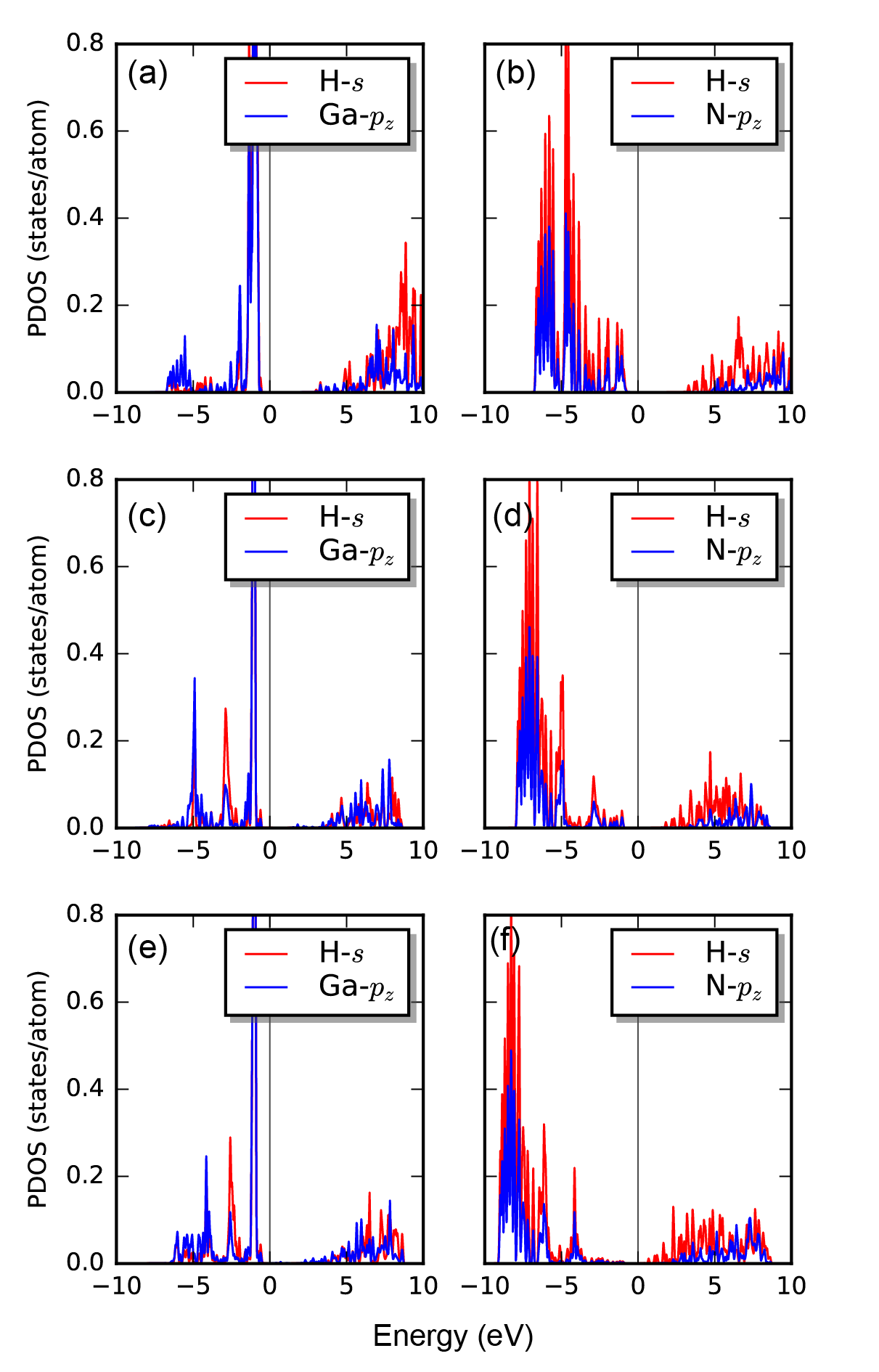}
\caption{\label{figs1}(Color online) PBE-calculated partial density of states of H-$s$, Ga-$p_z$ and N-$p_z$ in H-passivated monolayer [(a) and (b)], bilayer [(c) and (d)], trilayer [(e) and (f)] of GaN. }
\end{figure}

\begin{figure}[htbp]
\includegraphics[scale=1.5]{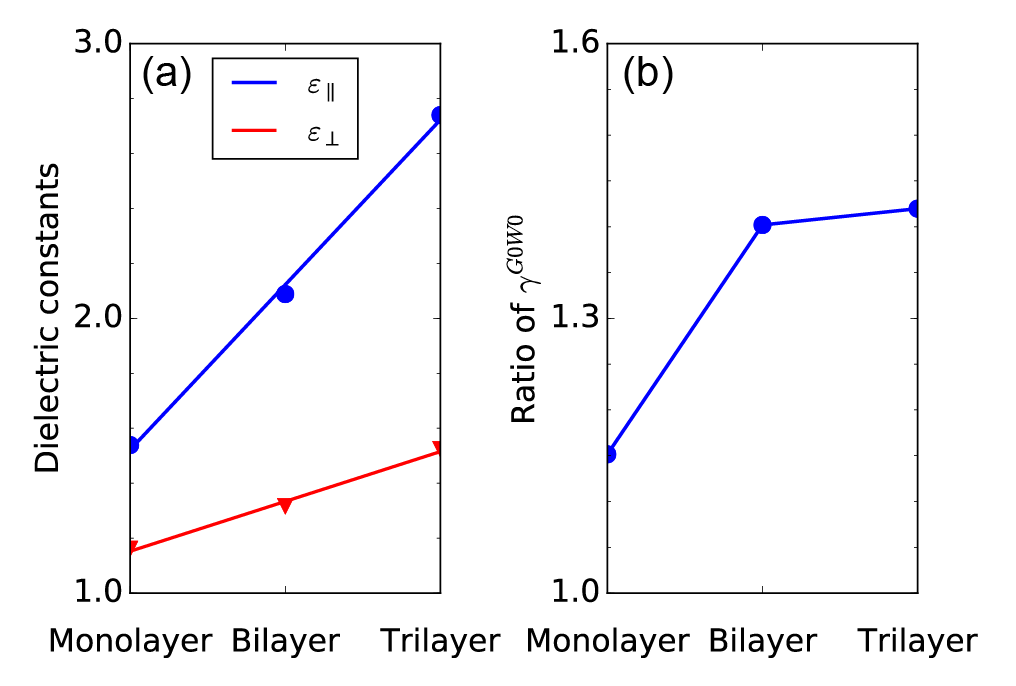}
\caption{\label{figs2}(Color online) PBE-calculated (a) static electronic dielectric constant along the parallel direction $\varepsilon_{\parallel}$ and perpendicular direction $\varepsilon_{\perp}$ to the layer, and (b) ratio of gap magnitude $\gamma^{G0W0}$ (for definition, see text) in buckled GaN film versus the layer number with a fixed supercell thickness of 23 {\AA}.}
\end{figure}

\begin{figure*}[htbp]
\includegraphics[scale=0.7]{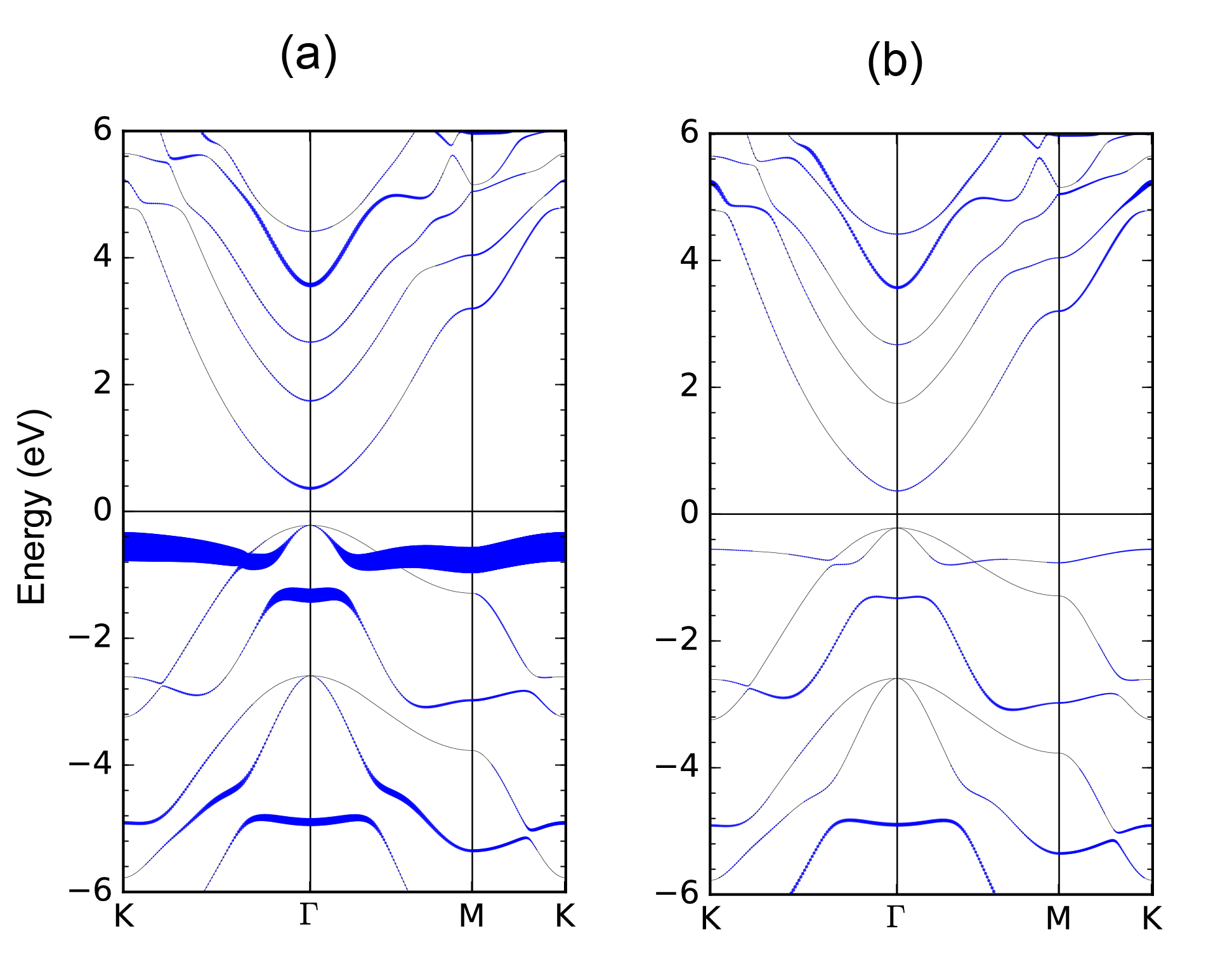}
\caption{\label{figs3}(Color online) PBE-calculated orbital-projected band structure of (a) H@Ga and (b) H@N atoms in the H-passivated buckled bilayer sheet. The line width indicates the weight of the component. The Fermi level is set to zero.}
\end{figure*}

\begin{figure*}[htbp]
\includegraphics[scale=0.7]{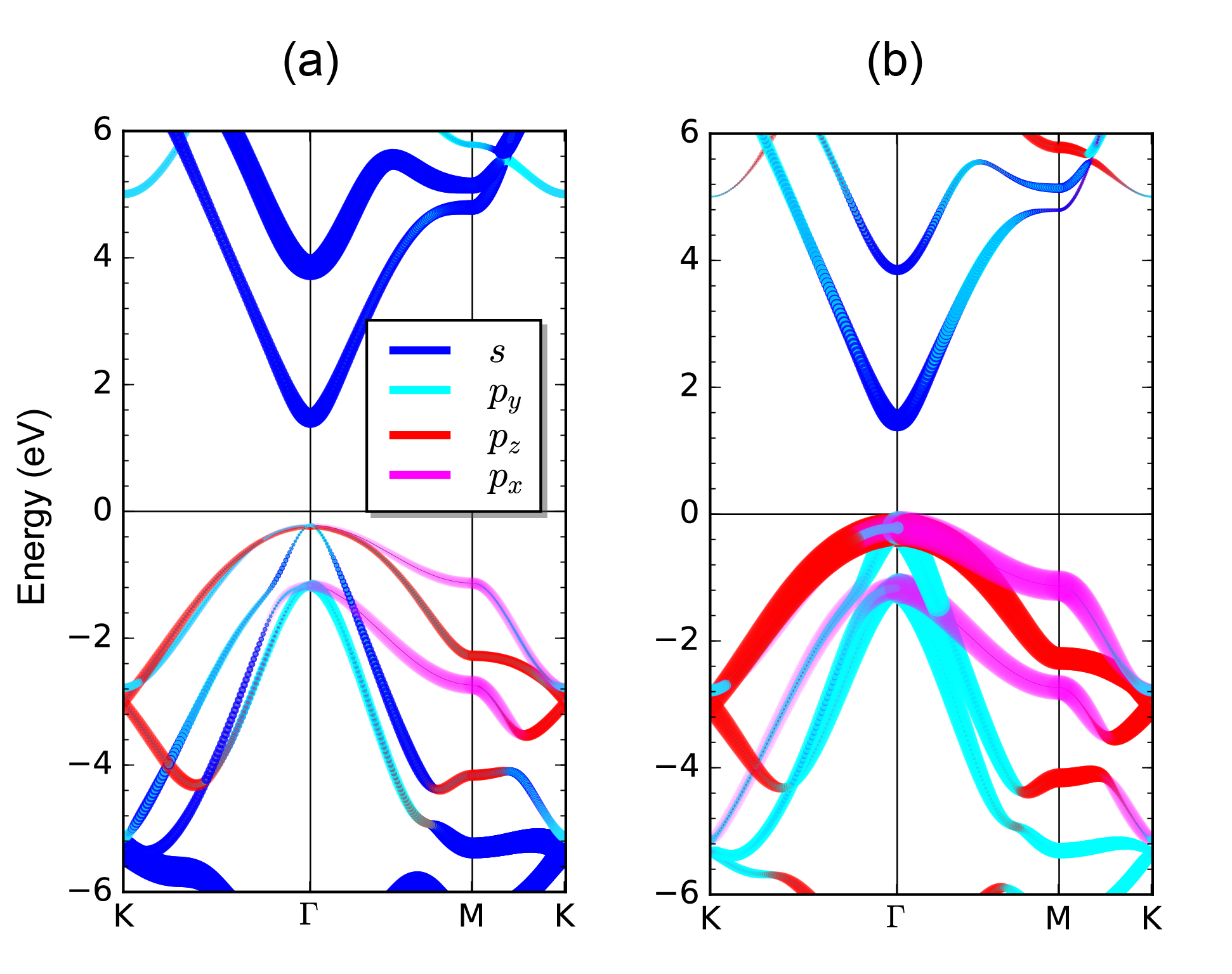}
\caption{\label{figs4}(Color online) PBE-calculated orbital-projected band structure of (a) Ga and (b) N atoms in bulk GaN. The line width indicates the weight of the component. The Fermi level is set to zero.}
\end{figure*}

\begin{figure*}[htbp]
\includegraphics[scale=1.6]{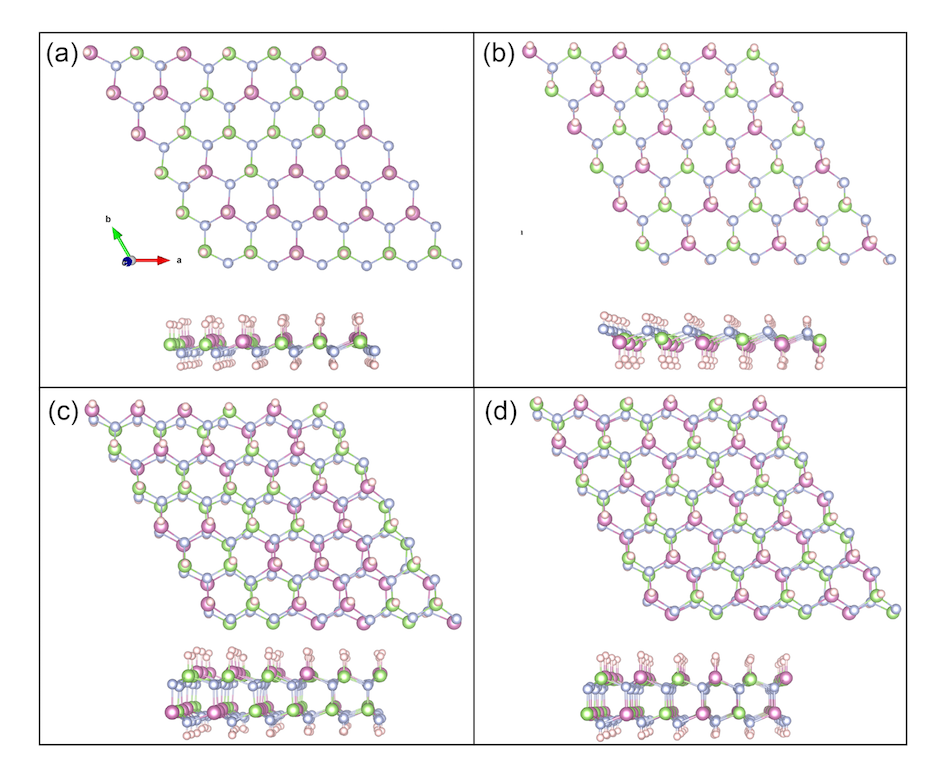}
\caption{\label{figs5}(Color online) Top and side views of the optimized structure of disordered  monolayer (a) and bilayer (b), ordered monolayer (c) and  bilayer (d) for the buckled In$_x$Ga$_{1-x}$N alloys. The green, silver and pink spheres refer to Ga, N and H atoms respectively.}
\end{figure*}

\begin{figure*}[htbp]
\includegraphics[scale=0.4]{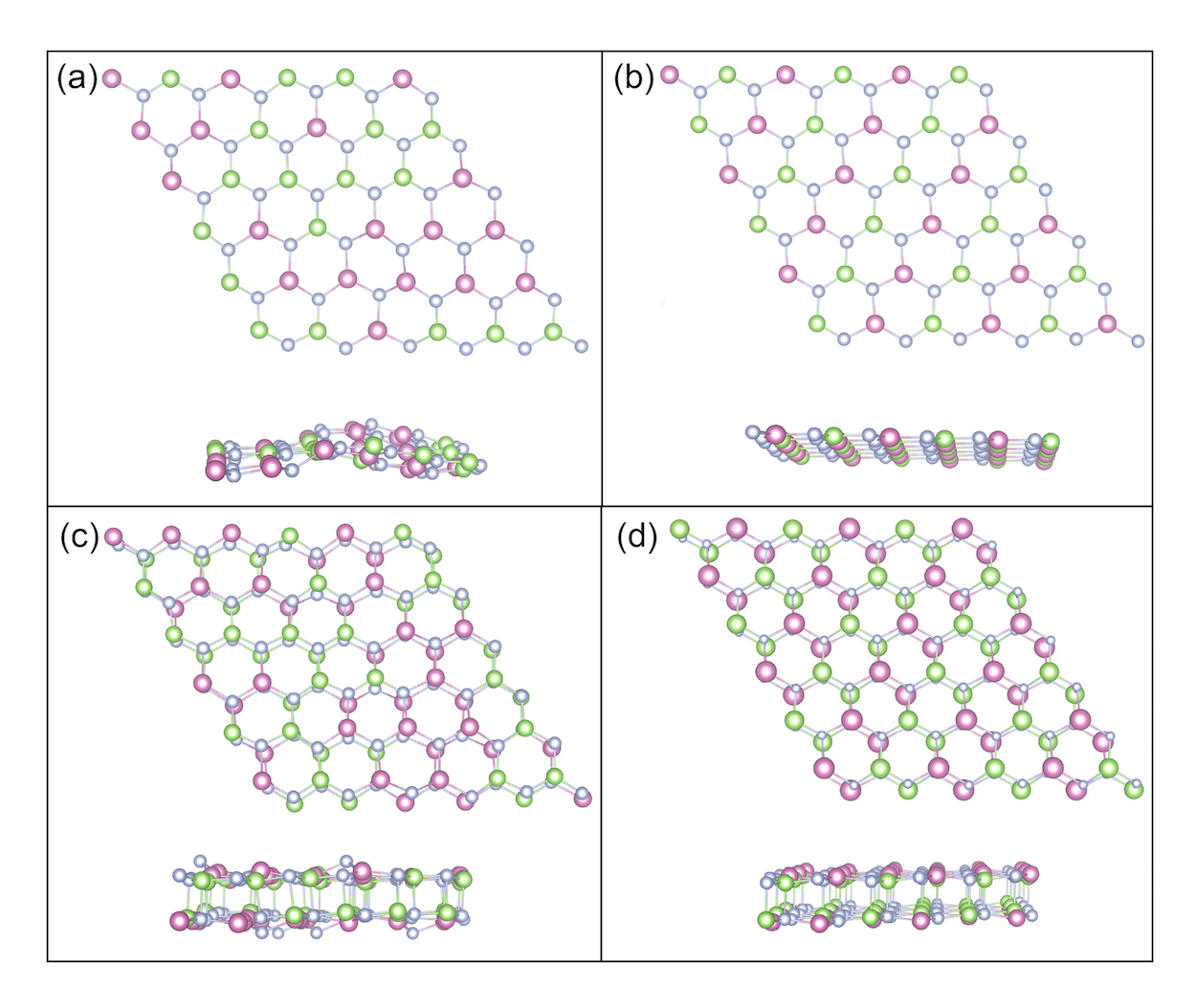}
\caption{\label{figs6}(Color online) Top and side views of the optimized structure of disordered  monolayer (a) and bilayer (b), ordered monolayer (c) and  bilayer (d) for the planar In$_x$Ga$_{1-x}$N alloys. The green, silver and pink spheres refer to Ga, N and H atoms respectively.}
\end{figure*}

\end{document}